\documentclass{emulateapj}
\usepackage[]{natbib, graphicx}

\begin{document}

\title{The sudden death of the nearest quasar}

\shorttitle{The sudden death of the nearest quasar}
\shortauthors{Schawinski et al.}
\slugcomment{To appear in the Astrophysical Journal Letters}

\author{
Kevin Schawinski,\altaffilmark{1,2,3}, 
Daniel A. Evans,\altaffilmark{4,5,6}, 
Shanil Virani,\altaffilmark{2,3,7}, 
C. Megan Urry,\altaffilmark{2,3,7}, 
William C. Keel,\altaffilmark{8,9}, 
Priyamvada Natarajan,\altaffilmark{2,3,7}, 
Chris J. Lintott,\altaffilmark{10,11},
Anna Manning,\altaffilmark{8,9}, 
Paolo Coppi,\altaffilmark{2,3,7}, 
Sugata Kaviraj,\altaffilmark{10, 12}, 
Steven P. Bamford,\altaffilmark{13},
Gyula I. G. J\'ozsa,\altaffilmark{14,15}, 
Michael Garrett,\altaffilmark{14,16,17}, 
Hanny van Arkel,\altaffilmark{14}, 
Pamela Gay,\altaffilmark{18}
and Lucy Fortson\altaffilmark{19}
}

 \altaffiltext{1}{Einstein Fellow}
 \altaffiltext{2}{Department of Physics, Yale University, New Haven, CT 06511, U.S.A.}
 \altaffiltext{3}{Yale Center for Astronomy and Astrophysics, Yale University, P.O. Box 208121, New Haven, CT 06520, U.S.A.}
 \altaffiltext{4}{Massachusetts Institute of Technology, Kavli Institute for Astrophysics and Space Research, 77 Massachusetts Avenue, Cambridge, MA 02139, USA}
 \altaffiltext{5}{Harvard-Smithsonian Center for Astrophysics, 60 Garden Street, Cambridge, MA 02138, USA}
 \altaffiltext{6}{Elon University, Elon, NC 27244, USA}
 \altaffiltext{7}{Department of Astronomy, Yale University, New Haven, CT 06511, USA}
 \altaffiltext{8}{Department of Physics and Astronomy, University of Alabama, Box 870324,Tuscaloosa, AL 35487, USA}
 \altaffiltext{9}{Visiting Astronomer, Kitt Peak National Observatory, NOAO, operated by AURA under cooperative agreement with the US NSF}
 \altaffiltext{10}{Astrophysics Department, University of Oxford, Oxford, OX1 3RH, UK}
 \altaffiltext{11}{Adler Planetarium, 1300 S. Lakeshore Drive, Chicago, IL 60605}
 \altaffiltext{12}{Blackett Laboratory, Imperial College London, London SW7 2AZ, UK}
 \altaffiltext{13}{Centre for Astronomy \& Particle Theory, University of Nottingham, University Park, Nottingham NG7 2RD, UK}
 \altaffiltext{14}{Netherlands Institute for Radio Astronomy, Postbus 2, 7990 AA Dwingeloo, The Netherlands }
 \altaffiltext{15}{Argelander-Institut f\"ur Astronomie, Auf dem H\"ugel 71, 53121 Bonn, Germany}
 \altaffiltext{16}{Leiden Observatory, University of Leiden, P.O. Box 9513, 2300 RA Leiden, The Netherlands}
 \altaffiltext{17}{Centre for Astrophysics and Supercomputing, Swinburne University of Technology, Hawthorn, 3122 Victoria, Australia}
 \altaffiltext{18}{Southern Illinois University Edwardsville, IL, USA}
 \altaffiltext{19}{School of Physics and Astronomy, 116 Church Street S.E., University of Minnesota/Twin Cities, Minneapolis, MN 55455, USA}

\email{kevin.schawinski@yale.edu}

\def\Chandra{\textit{Chandra}}
\def\XMM{\textit{XMM-Newton}}
\def\Swift{\textit{Swift}}
\def\Suzaku{\textit{Suzaku}}

\def\Ebmv{E($B-V$)}
\def\Lbol{$L_{\rm bol}$}
\def\LOIII{$L[\mbox{O\,{\sc iii}}]$}
\def\Ledd{${L_{\rm Edd}}$}
\def\LLedd{${L/L_{\rm Edd}}$}
\def\LOIIIs4{$L[\mbox{O\,{\sc iii}}]$/$\sigma^4$}
\def\LOIIIMbh{$L[\mbox{O\,{\sc iii}}]$/$M_{\rm BH}$}
\def\Mbh{$M_{\rm BH}$}
\def\Msigma{$M_{\rm BH} - \sigma$}
\def\Ms{$M_{\rm *}$}
\def\Msun{$M_{\odot}$}
\def\Msunyr{$M_{\odot}yr^{-1}$}
\def\HI{H{\sc i}}

\def\ergs{$~\rm erg s^{-1}$}
\def\kms{$~\rm km s^{-1}$}
\def\pcms{$~\rm cm^{-2}$}

\def\OI{[\mbox{O\,{\sc i}}]~$\lambda 6300$}
\def\OIII{[\mbox{O\,{\sc iii}}]~$\lambda 5007$}
\def\SII{[\mbox{S\,{\sc ii}}]~$\lambda \lambda 6717,6731$}
\def\NII{[\mbox{N\,{\sc ii}}]~$\lambda 6584$}

\begin{abstract}
Galaxy formation is significantly modulated by energy output from supermassive black holes at the centers of galaxies which grow in highly efficient luminous quasar phases. The timescale on which black holes transition into and out of such phases is, however, unknown. We present the first measurement of the shutdown timescale for an individual quasar using X-ray observations of the nearby galaxy IC 2497, which hosted a luminous quasar no more than 70,000 years ago that is still seen as a light echo in `Hanny's Voorwerp', but whose present-day radiative output is lower by at least 2 and more likely by over 4 orders of magnitude. This extremely rapid shutdown provides new insights into the physics of accretion in supermassive black holes, and may signal a transition of the accretion disk to a radiatively inefficient state.
\end{abstract}


\begin{figure*}

\begin{center}
\includegraphics[width=0.66\textwidth]{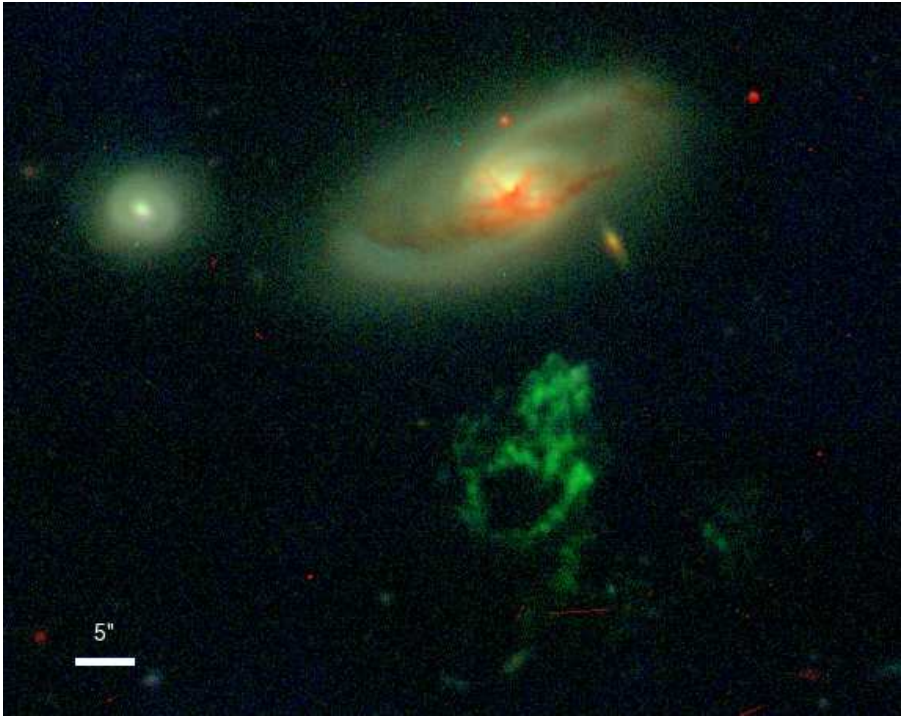}
\end{center}

\caption{Ground-based optical image from WIYN of IC 2497 (top), Hanny's Voorwerp (bottom) and a nearby companion galaxy (left). This image is a composite of $B$, $V$ and $I$ band filters taken with the WIYN telescope in excellent $\sim0.''45$ seeing. Since the Voorwerp is a pure emission line object dominated by a powerful \OIII\ line, it only appears in the $V$ band image (green). The bar in the lower-left indicates a scale of $5''$, which at the redshift of IC 2497 corresponds to just under 5 kpc. To ionize and light up the Voorwerp, a quasar point source in the nucleus of IC 2497 has to have a bolometric luminosity of at least $L_{\rm bol} \sim 10^{46}$\ergs, but no point source indicating the presence of such a luminous quasar is evident. The prominent dust lanes in the bulge of IC 2497 indicate an ongoing morphological disturbance that may be related to the nearby galaxy to the left or to a recent merger. \label{fig:image}}

\end{figure*}

\section{Introduction}
\label{sec:intro}

The discovery of the object known as ``Hanny's Voorwerp''\footnote{Voorwerp is the Dutch word for object. The object's name was coined by the members of the Galaxy Zoo forum who named it after the discoverer, Hanny van Arkel (\texttt{http://www.galaxyzooforum.org}).} (the Voorwerp hereafter) by a citizen scientist participating in the Galaxy Zoo project \citep{2008MNRAS.389.1179L, 2009MNRAS.399..129L} permits the first direct probe of a quasar's variability for an individual source on timescales significantly longer than human lifetimes \citep{2009MNRAS.399..129L}. The Voorwerp is a large ($11 \times 16$ kiloparsec) cloud of ionized gas 45,000-70,000 lightyears away from the nucleus of the galaxy IC 2497, embedded in a larger reservoir of atomic hydrogen \citep{2009A&A...500L..33J}, with a mass of several times $10^{9}$\Msun\ (see Figure \ref{fig:image}). The optical spectrum of the Voorwerp is dominated by a powerful \OIII\ emission line and shows little detectable continuum. The presence of other emission lines with high ionization potentials, such as $[\mbox{He\,{\sc ii}}]$ and [\mbox{Ne\,{\sc v}}], together with the narrowness of the emission lines suggest that the Voorwerp is being photoionized by the hard continuum of an active galactic nucleus (AGN; an accreting supermassive black hole), rather than by other processes such as star formation or shocks (such as might be induced by a jet; \citealt{2009MNRAS.399..129L}). Radio observations of IC 2497 reveal a nuclear source and a jet hotspot in the nucleus, and a large kiloparsec-scale structure that may be a jet \citep{2009A&A...500L..33J, 2010A&A...517L...8R}. The Voorwerp lies where this jet meets the \HI\  reservoir and coincides with a local decrement in atomic hydrogen presumably due to photoionization (see Fig. 2 of ref. \citealt{2009A&A...500L..33J}). An actively accreting black hole at the centre of IC 2497 is therefore the only plausible source of ionization that can account for the emission seen from the Voorwerp.

The light travel time from the nucleus of IC 2497 to the Voorwerp, accounting for all possible geometries, ranges from 45,000 to 70,000 years \citep{2009MNRAS.399..129L} and so the Voorwerp reflects the radiative output of the central black hole of IC2497 at those times in the past. Its large physical extent rules out a brief flare, caused for example by the tidal disruption of a star, as the ionizing source \citep{2004ApJ...603L..17K, 2006ApJ...653L..25G}. The black hole must produce sufficient ionizing photons in order to power its observed \OIII-luminosity. This requirement corresponds to $2 \times 10^{45}$\ergs\ between 1 and 4 Rydberg (13.6-54.4 eV), assuming isotropic emission. This likely underestimates the necessary intrinsic luminosity due to large amounts of dust obscuration in the bulge of IC 2497, which has prominent dust lanes (Figure \ref{fig:image}). We calculate the required luminosity of the quasar lighting up the Voorwerp by taking a template spectral energy distribution (SED) for an unobscured quasar from \cite{1994ApJS...95....1E} which the Voorwerp is presumably seeing. We then scale this template to match the minimum UV ionizing luminosity needed and derive a minimum bolometric luminosity of $L_{\rm bol, past} = 1.2\times10^{46}$\ergs. This means IC 2497, at a redshift of $z= 0.0502$ \citep{1995ApJS..100...69F}, is, or has been, the nearest luminous quasar, an extremely rare object in the local Universe.

However, IC 2497 poses a challenge: the optical image reveals no strong point source (Figure \ref{fig:image}), the nuclear spectrum shows very weak optical line emission \citep{2009MNRAS.399..129L}, and it also has a weak ($\sim 10^{38}$\ergs) nuclear radio source \citep{2009A&A...500L..33J}. These observations are difficult to reconcile with the presence of a currently active $L_{\rm bol} \sim 10^{46}$\ergs\ quasar. There are two possible scenarios that can account for these apparently contradictory observations as argued by \cite{2009MNRAS.399..129L}: 1) the quasar in IC 2497 features a novel geometry of obscuring material and is obscured at an unprecedented level only along our line of sight, while being virtually unobscured towards the Voorwerp; or 2) the quasar in IC 2497 has shut down within the last 70,000 years, while the Voorwerp remains lit up due to the light travel time from the nucleus. If the latter is the case, the IC 2497--Voorwerp system gives for the first time an upper limit of the shutdown timescale of an individual quasar central engine. In this Letter, we present observations to distinguish these two scenarios.

\section{Observations and Results}
\label{sec:obs}

\subsection{Archival Infrared Data}
We have obtained multiple, independent observations to assess the current nuclear luminosity of IC 2497. Light from an obscured $L_{\rm bol} \sim 10^{46}$\ergs\ quasar in IC 2497 should be re-emitted at mid- and far-infrared wavelengths. Archival IRAS observations show that the infrared temperature of IC 2497 of $f_{25\mu\rm{m}}/f_{60\mu\rm{m}} = 0.1$ is cold compared to that of nearby, luminous highly-obscured AGN \citep{1990IRASF.C......0M, 1999ApJ...522..157R} and IC 2497 also adheres to the radio-far infrared correlation \citep{2010A&A...517L...8R}. Thus, IC 2497 does not exhibit any evidence for reprocessed infrared emission from an active nucleus.

\subsection{Suzaku and XMM-Newton Observations}
The presence of a quasar with high levels of obscuration should still be detected in the hard X-rays ($>10$ keV) where photoelectric absorption is minimal. To this end, we observed IC 2497 with the \Suzaku\ X-ray space observatory for 75 kiloseconds on 2009-04-20 using both the XIS (0.2--12 keV) and HXD/PIN (10--600 keV) detectors. This observation was designed to be sufficiently deep to detect a quasar with $L_{\rm bol} \sim 10^{46}$\ergs\ and an obscuring column of $N_{\rm H} = 10^{24}$\pcms. There is no significant detection with the XHD/PIN instrument consistent with such a highly obscured quasar, although there are some marginally significant counts at E $> 15$ keV; even if real, these imply a 10--20 keV luminosity orders of magnitude below the required luminosity.

\begin{figure}

\begin{center}
\includegraphics[ width=0.49\textwidth]{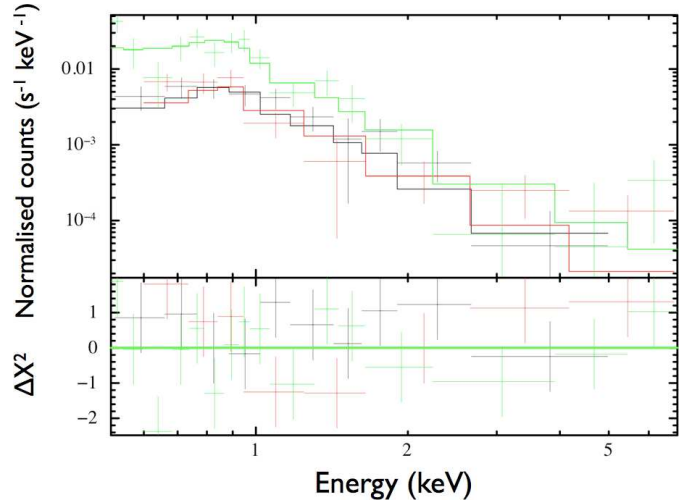}
\end{center}

\caption{X-ray spectrum from \XMM\ from the EPIC-pn (black), MOS1 (red) and MOS2 (green). The solid lines  are the combined model of a power law due to an unobscured AGN with a present-day luminosity of $L_{2-10 ~ \rm{keV}} = 4.2 \times 10^{40}$\ergs, and a collisionally ionized plasma from hot, diffuse gas. Residuals from the best-fit two-component model (see text) are shown in the bottom panel. Extrapolated to hard X-rays, the fit to the power law is consistent with the marginal PIN counts at $\sim$15 keV.\label{fig:data}}

\end{figure}

We then obtained a second observation of IC 2497 with the \XMM\ X-ray space observatory using the EPIC-pn, MOS-1 and MOS-2 detectors on 2010-04-19 with a total useful observing time of 11 kiloseconds and sensitivity between 0.1 and 7 keV. With \XMM, we detect a source at 0.1-5 keV, which can be fit with a combination of two spectral components: a collisionally ionized plasma ($T=0.78^{+0.18}_{-0.14}$ keV), consistent with thermal emission from a warm interstellar medium in IC 2497, and an unabsorbed power law ($\Gamma = 2.5\pm0.7$) from a very low luminosity AGN with $L_{2-10 ~\rm{keV}} = 4.2 \times 10^{40}$\ergs. The emission may equally well be attributed to emission from star formation and X-ray binaries. Neither the \XMM, nor the \Suzaku\ XIS observations detect the K$\alpha$ line feature at $\sim$6.4 keV which is prominent in obscured AGN, especially the most highly obscured systems \citep{2007ApJ...664L..79U}. We show the \XMM\ spectra in Figure \ref{fig:data}. If the observed X-ray power law is not due to a low-luminosity AGN, then its radiative output must be even lower. 

If we assume that all the observed emission is from non-AGN sources, then the \Suzaku\ data can give us an extreme upper limit on the present-day AGN luminosity. Assuming a Compton-thick AGN, the full \XMM\ and \Suzaku\ data, especially the PIN data, limit the present-day hard X-ray luminosity to $L_{15-30 ~\rm{keV}} = 3.5 \times 10^{42}$\ergs, roughly two orders of magnitude more luminous than the observed soft X-ray power law, but not sufficient to ionize the Voorwerp. The interpretation of the power law seen by \XMM\ as a low-luminosity AGN appears to be the more likely one as we see a low-luminosity AGN in two other wavelength regimes: a VLBI radio core \citep{2010A&A...517L...8R} and a nuclear point source visible in a \textit{Hubble Space Telescope} F184W image \citep{keel_hst} indicate that a low-luminosity AGN consistent with the observed X-ray power law is present in IC 2497.

The \XMM\ detection of a power law continuum is consistent with an active black hole at the centre of IC 2497 but at a very low accretion rate. It must also be unobscured along our line of sight, apart from Galactic extinction ($N_{\rm H} = 1.31 \times10^{20}$\pcms). Additional absorption does not improve the quality of the fit, and the amount of absorption is consistent with zero. This power-law AGN is therefore a fair probe of the current radiative output of the central engine of IC 2497. An extrapolation of the fit to the \XMM\ data out to the energies sampled by \Suzaku\ PIN is consistent with the marginal \Suzaku\ PIN detection. 

The hypothesis that a quasar of the required luminosity is present, but sufficiently obscured along our line of sight to elude detection at optical wavelengths is ruled out by the \Suzaku\ hard X-ray luminosity limit and the lack of a K$\alpha$ line at 6.4 keV. The \Suzaku\ upper limit implies a drop in luminosity of the quasar in IC 2497 of at least 2 orders of magnitude. The more likely scenario is that the power law component of the soft X-ray spectrum is that of the present-day low-luminosity AGN state of the central engine. Using the ionizing quasar SED template scaled to the luminosity required by the Voorwerp, the expected X-ray luminosity in this energy band would be  $L_{2-10 ~\rm{keV}} = 8 \times 10^{44}$\ergs, while we only detect $4.2 \times 10^{40}$\ergs, a discrepancy of over 4 orders magnitude.

\section{Discussion}
\label{sec:discussion}

The low X-ray luminosity definitively rules out the presence of a currently active quasar in IC2497, and we therefore conclude that the galaxy's central engine has decreased its radiative output by at least 2 and more likely by over 4 orders of magnitude since being in a much more luminous phase  within the last 70,000 years. We have therefore for the first time constrained the shutdown timescale of an individual quasar. 

The sudden death of the quasar in IC 2497 might be due to a sharp decrease in the fuel supply or a change of accretion state. Rapid changes between `high' (radiatively efficient) and `low-hard' (radiatively inefficient) states most likely driven by instabilities in the accretion disk have been seen in Galactic X-ray binaries (XRBs) which typically have black holes of a few solar masses \citep{2000ApJ...535..798N, 2004ARA&A..42..317F,2004MNRAS.355.1105F, 2010arXiv1005.3304P}. It has been suggested that supermassive black holes are essentially scaled-up versions of XRBs with similar black hole accretion physics \citep[e.g.][]{2003MNRAS.345L..19M,2006Natur.444..730M, 2006MNRAS.372.1366K}. IC 2497 may be the first quasar where such a rapid transition between accretion states is observed, of the kind routinely seen in XRBs.

The immediate question this analogy raises is whether the time scale of the state change corresponds to what we would expect from XRBs. State changes in the Galactic XRB GRS 1915+105, with a black hole mass of 10 \Msun, are detected on timescales of 1 hour. Scaling up linearly to the case of IC 2497 with a $10^9$ \Msun\ black hole (estimated from its bulge mass; \citealt{2004ApJ...604L..89H}) yields a timescale of 10,000 years. A more detailed calculation taking into account the ratio of bolometric luminosity to the Eddington luminosity \citep{2005MNRAS.364..208D} yields a timescale on the order of 10,000--100,000 years. Both are broadly consistent with our upper limit on the shutdown timescale. 

This concordance of timescales supports the interpretation of IC 2497 as a system that has transitioned from a classical quasar in a high state to a radiatively inefficient state where the bulk of the energy is dissipated not as radiation but as either thermal or kinetic energy \citep{1984RvMP...56..255B, 1994ApJ...428L..13N}. The presence of a recent radio outflow \citep{2010A&A...517L...8R} extending over $\sim 1000$ light years (projected distance) from the nucleus of IC 2497 also supports the hypothesis of a change in accretion state \citep{1995Natur.374..623N} if the launch of the jet is associated with the state change. 

However, the four orders of magnitude drop in luminosity poses a problem for a direct analogy. In XRBs, such large luminosity changes are seen as they return to quiescence after the (quasi-exponential) outburst decline. Typically it takes a few days for the cooling wave to propagate through to the inner disc in black hole binaries, as seen in observations \cite[e.g.,][]{1997ApJ...491..312C} and in theoretical lightcurves from disc instability models \citep[e.g.][]{2001A&A...373..251D}. Taking 1 day as the transition time scale and scaling from 10 \Msun\ to $10^{9}$ \Msun\ using the linear scaling yields a transition time scale for the quasar of at least 280,000 years, significantly longer that what we see in IC 2497. A change of accretion state remains a possible explanation for the observed luminosity drop in IC 2497, but in that case the analogy to XRBs does not scale linearly with the black hole mass. We therefore conclude that the sudden death of IC 2497 is a vital clue to how quasars accrete and shut down, but that we do not yet understand the physics of this process. 

If such a change of state in the accretion disk did occur, then it is entirely plausible that the accretion disk may change back to a high luminosity state on a similar timescale. Future multi-wavelength monitoring of IC 2497 could reveal such a change. Since the quasar in IC 2497 shut down less than 70,000 years ago, it offers an unobstructed view of the host galaxy of a quasar. The close distance of IC 2497 furthermore means that we can view this quasar host galaxy in greater detail than any other system. As such, it is ideally suited for observationally probing the fueling of the black hole and how the quasar phase is affecting the large-scale environment of the host galaxy, and in particular, whether it retains any evidence for whether the central engine was, or currently is, injecting kinetic or thermal energy into the interstellar medium and therefore doing feedback work.

\acknowledgements 

We thank the anonymous referee for helpful comments. This work is based on observations with the \XMM\ and \Suzaku\ X-ray satellites and the WIYN observatory, and was supported by NASA grants NNX09AR22G and NXX09AV69G. Support for the work of KS was provided by NASA through Einstein Postdoctoral Fellowship grant number PF9-00069 issued by the Chandra X-ray Observatory Center, which is operated by the Smithsonian Astrophysical Observatory for and on behalf of NASA under contract NAS8-03060. PN acknowledges the award of a Guggenheim fellowship. CJL acknowledges support from The Leverhulme Trust and a STFC Science and Society Fellowship. We thank Charles Bailyn, Chris Done and Phil Hopkins for discussions and suggestions.

This research has made use of NASA's Astrophysics Data System Bibliographic Services. \\
{\it Facility:} \facility{XMM-Newton (EPIC-pn, MOS)}, \facility{Suzaku (XIS, PIN), \facility{WIYN}}

\bibliographystyle{apj}


\end{document}